\documentclass[pre,twocolumn,superscriptaddress,showpacs,floatfix]{revtex4}
\usepackage{bm}
\usepackage{amsmath}
\usepackage{amssymb}
\usepackage{amsbsy}
\usepackage{latexsym}
\usepackage{multirow}
\usepackage{graphics}
\usepackage{graphicx}
\usepackage{epsfig}
\usepackage{color}
\definecolor{r}{rgb}{1,0,0}

\newcommand{\JHU}{Department of Mechanical Engineering\\
Institute for Data Intensive Engineering and Science\\
Johns Hopkins University, Baltimore, MD 21218}

\begin{document}

\title{Lagrangian Refined Kolmogorov Similarity Hypothesis for Gradient Time-evolution in Turbulent Flows}

\author{Huidan Yu}
\email{hyu36@jhu.edu}

\author{Charles Meneveau}
\email{meneveau@jhu.edu}
\affiliation{\JHU}


\begin{abstract}

We study the time evolution of velocity and pressure gradients in isotropic turbulence, by quantifying their decorrelation time scales as one follows fluid particles in the flow. The Lagrangian analysis uses data in a public database generated using direct numerical simulation of the Naiver-Stokes equations, at a Reynolds number $Re_{\lambda} \approx 430$. It is confirmed that when averaging over the entire domain, correlation functions decay on timescales on the order of the mean Kolmogorov turnover time scale, computed from the globally averaged rate of dissipation and viscosity.  However, when performing the analysis in different subregions of the flow,  turbulence intermittency leads to large spatial variability in the  decay time scales. Remarkably, excellent collapse of the auto-correlation functions is recovered when using the `local Kolmogorov time-scale' defined using the locally averaged, rather than the global, dissipation-rate. This provides new evidence for the validity of Kolmogorov's Refined Similarity Hypothesis, but from a Lagrangian viewpoint that provides a natural frame to describe the dynamical time evolution of turbulence.

\end{abstract}

\maketitle
Understanding the universal features of the dynamics of turbulence \cite{rf:Kolmogorov41,rf:Frisch95} continues to be a formidable problem in classical and statistical physics. The original Kolmogorov 1941 theory relied on the overall averaged dissipation-rate $\langle \epsilon \rangle$ to predict, among others, the scaling properties of the energy spectrum. It was extended to account for intermittency by the introduction of the  refined Kolmogorov similarity hypothesis (RKSH) \cite{rf:Kolmogorov62}. In this extension, attention is placed on conditional statistics based on the locally averaged dissipation-rate in some particular subregion of the flow, such as a sphere or box of size $r$. This local dissipation-rate, usually denoted by $\epsilon_r$, is defined according to
\begin{equation}
 \epsilon_r({\bf x}) = \frac{1}{V}\int \limits_{{\cal{R}}_r({\bf x})} 2 \nu \left[S_{ij}({\bf x}')\right]^2 d^3 {\bf x}',
 \label{eq:defepsilonr}
 \end{equation}
where $V$ is the volume of the subregion ${\cal{R}}_r({\bf x})$ of size $r$ centered at ${\bf x}$, $\nu$ is the kinematic viscosity of the fluid, and $S_{ij}$ is the strain-rate tensor defined as $S_{ij}=(\partial u_i / \partial x_j+\partial u_j / \partial x_i)/2$ (where $u_i$ is the velocity field).  The longitudinal velocity increment at scale $r$ is defined as $\delta_r u = [u_i({\bf x} + {\bf r}) - u_i({\bf x})](r_i/r)$, and the RKSH states that in the inertial range of turbulence the statistics of $\delta_r u$ depend on $r$ and $\epsilon_r$ so that from dimensional analysis moments of $\delta_r u$, conditioned upon a fixed value of $\epsilon_r$ will scale as $\langle \delta_r u^p \vert \epsilon_r \rangle = C_p (r \epsilon_r )^{p/3}$ according to Kolmogorov's 1941 postulate. Anomalous scaling then results from the additional global averaging and anomalous scaling behavior of moments of  $\epsilon_r$.

Extensive literature  to test RKSH has focused mainly on velocity increments \cite{rf:Stolovitzky92,rf:Thoroddsen,rf:Chen93,rf:Stolovitzky94,rf:Chen95,rf:Ching08} or acceleration \cite{rf:Yeung06}. In recent years there has been growing attention placed in the dynamical evolution of the velocity gradient tensor ${\bf A}$ ($A_{ij}\equiv \partial u_i / \partial x_j$) due to the fact that ${\bf A}$ provides rich information about the topological and statistical properties of small-scale structure in turbulence. The Lagrangian time evolution of ${\bf A}$ can be obtained by taking gradient of the NS equation \cite{rf:Vieillefosse82}:
\begin{equation}
\frac{dA_{ij}}{dt}=-A_{ik}A_{kj}-\frac{\partial ^2p}{\partial x_i \partial x_j}+\nu \frac{\partial ^2A_{ij}}{\partial x_k \partial x_k}
\label{eq:Aij}
\end{equation}
where $d/dt$ stands for Lagrangian material derivative and $p$ is the pressure divided by the density of the fluid. The first term on the right-hand side of Eq. (\ref{eq:Aij}) denotes the nonlinear self-interaction of $\bf A$, the second term is a tensor called pressure Hessian $P_{ij}\equiv \partial ^2p/\partial x_i \partial x_j$, and the third is the viscous term. Assuming the pressure Hessian is isotropic (i.e.neglecting $\partial^2_{ij}p-\partial^2_{kk}p ~\delta_{ij}/3$) and neglecting the viscous term lead to a closed formulation for $\bf A$, the so-called Restricted-Euler (RE) equation, which has analytical solutions for the full tensor-level time history. Remarkably this simple system is already sufficient to explain a number of non-trivial geometrical trends found in real turbulence \cite{rf:Vieillefosse82,rf:Cantwell92}. Nevertheless, the RE system leads to nonphysical finite-time singularities because the self-stretching is not constrained by any energy exchange or loss mechanism in the system. To develop models for such energy exchange mechanisms, there is a need to better understand the time evolution of $\bf A$, as one follows fluid particles across a turbulent flow. As will be shown in this paper, the RKSH can play a crucial role in determining the characteristic time scales of this evolution in different parts of the flow.

 Much effort at regularizing the RE system to avoid the nonphysical singularity has been made \cite{rf:Girimaji90,rf:Martin98,rf:Chertkov99,rf:Jeong03,rf:Chevillard06,rf:Biferale07} in the past two decades. Despite progress, the inability to fully account for the anisotropic pressure Hessian and viscous effects continues to limit the accuracy of the existing models for the evolution of velocity-gradient tensor \cite{rf:Afonso09}.  In various models of the Lagrangian dynamics \cite{rf:Girimaji90,rf:Jeong03,rf:Chevillard06,rf:Afonso09}, the characteristic correlation times along Lagrangian trajectories and their scaling with Reynolds number play a central role. For instance in the model based on the `recent fluid deformation closure' \cite{rf:Chevillard06} the Lagrangian pressure Hessian tensor is assumed isotropic, based on the idea that any causal relationship between initial and present orientations will be lost after a characteristic Lagrangian correlation time-scale of the tensor  ${\bf A}$.  The usual expectation is that the characteristic correlation time-scale of ${\bf A}$ is the Kolmogorov time-scale $\tau_K = (\nu/\langle \epsilon \rangle)^{1/2}$.

Since the dynamical equation for the velocity gradient tensor ${\bf A}$ (Eq. \ref{eq:Aij})  and the existing models \cite{rf:Girimaji90,rf:Martin98,rf:Chertkov99,rf:Jeong03,rf:Chevillard06,rf:Biferale07} are written for the full tensor, it is of interest to quantify the temporal correlation function of
each tensor element but to do so in a fashion that is coordinate system invariant. We thus use the tensor-based Lagrangian time correlation function \cite{rf:Yu09} of a second-rank tensor ${\bf C}$, defined as
\begin{equation}
 \rho_{{\bf C}}(\tau) \equiv \frac{\langle{C_{ij}(t_0)C_{ij}(t_0+\tau)}\rangle}
 {\sqrt{\langle{(C_{mn}(t_0))^2}\rangle \cdot \langle{(C_{pq}(t_0+\tau))^2}\rangle}},
 \label{eq:correlationfunction}
\end{equation}
where $\tau$ is the time-lag along Lagrangian trajectories and $\langle{\cdots}\rangle$ may represent ensemble or volume averaging for homogeneous turbulence. Here tensor elements are assumed to have zero mean for isotropic turbulence. For a statistically steady-state process, when $\rho_{\bf C}(\tau)$ does not depend upon $t_0$, averaging can also be done over $t_0$. The correlation function $\rho_{\bf C}(\tau)$ provides a frame-invariant description of the auto-correlation structure of tensor elements, appropriately summed over all directions.

The fluid acceleration is another variable of great interest. The Lagrangian velocity and acceleration statistics have been investigated both experimentally \cite{rf:LaPorta01} and numerically \cite{rf:Yeung02,rf:Toschi09}. The acceleration is associated more closely with inertial-range and small-scale structures. It has been  established that the acceleration is dominated by the pressure gradient $\nabla p$ and viscous forces are negligible away from boundaries \cite{rf:Vedula99,rf:Liu06}. Therefore, we study the time correlations associated with the pressure gradient. Specifically, similar to the tensor correlation function defined in Eq. (\ref{eq:correlationfunction}), for the pressure gradient correlation $ \rho_{{\nabla}p}(\tau)$ we use the inner product of the vector at $t_0$ and $t_0+\tau$.

We measure $\rho_{\bf C}(\tau)$ for ${\bf C}={\bf A}$, ${\bf S}$ or ${\bf \Omega} (\Omega_{ij}=(A_{ij}-A_{ji})/2)$ and $\rho_{\nabla p}(\tau)$ using data from pseudo-spectral direct numerical simulation of forced isotropic turbulence on a $1024^3$-node periodic domain, with $Re_\lambda=433$. The 27 Terabytes of data are stored in a public database format (see {\verb http://turbulence.pha.jhu.edu) and can be accessed using web-service tools. We track fluid particles and extract Lagrangian information along the trajectories, such as velocity and pressure gradients through the second-order Runge-Kutta particle-tracking algorithm \cite{rf:Yeung88}. The interpolation of required velocities uses $8th$-order Lagrange polynomials in space and piecewise cubic Hermite polynomial interpolation in time, as implemented in the predefined functions in the database \cite{rf:Li08}.

The Lagrangian time correlations for {\bf S} and {\bf $\Omega$} are shown as open squares in Fig. \ref{Fig_Epsilonsort_SO_tensor}(a) (the meaning of the other lines are explained below). As it can be seen, the strain-rate auto-correlation decays fairly rapidly and reaches almost zero near 5-6 Kolmogorov timescales, much faster than that of the rotation-rate \cite{rf:Yu09}.

\begin{figure}[htbp]
   \begin{center}
      \includegraphics[width=3.2 in] {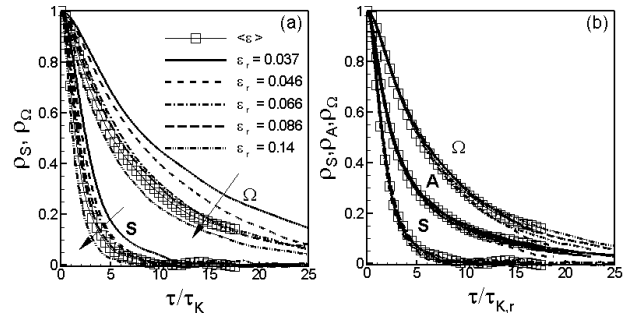}
      \caption{Lagrangian auto-correlations of strain- and rotation-rate tensors computed from Eq. (\ref{eq:correlationfunction}). Open squares are for global average over randomly located particles, whereas different lines correspond to subregions of the flow characterized by certain $\epsilon_r$. Time-lag is normalized using (a) the global Kolmogorov time scale $\tau_{K}$ and (b) the local time-scale $\tau_{K,r}$.}
      \label{Fig_Epsilonsort_SO_tensor}
   \end{center}
\end{figure}

It is known\cite{rf:Frisch95} that turbulence is highly intermittent, with regions displaying strong fluctuations, interspersed with less turbulent regions. In order to study intermittency from the viewpoint of the Lagrangian time evolution, we compute the correlation functions based on fluid particles that originate from various subregions of the flow. The subregions are characterized by the local dissipation-rate $\epsilon_r$. We thus define the conditional time-correlation functions based on $\epsilon_r$. In Eq.(\ref{eq:correlationfunction}), the three global averages are replaced by conditional averages, e.g. $\langle{C_{ij}(t_0)C_{ij}(t_0+\tau)}\rangle \to \langle{C_{ij}(t_0)C_{ij}(t_0+\tau)}\vert \epsilon_r \rangle$. In these averages the initial position of particles contributing to the average at time $t_0$ are sampled from several local boxes of size $r$ that have a prescribed locally averaged dissipation-rate $\epsilon_r$ (in practice a range of values in a prescribed `bin' is used). We consider four scales $r$ in the inertial range of turbulence corresponding to 16, 32, 64 and 128 grid-points of the DNS or $r=34\eta_K$, $68\eta_K$, $136\eta_K$ and $276\eta_K$ respectively where $\eta_K$ is the Kolmogorov length scale.  Four bins of $\epsilon_r$ values are chosen, centered at: $\epsilon_r = $0.0076, 0.0186, 0.0924 and 0.148 for $16$-cubes, 0.0096, 0.0186, 0.092 and 0.148 for $32$-cubes, and 0.0338, 0.0519, 0.0936 and 0.122 for $128$-cubes. For scale $r=136\eta_K$ ($64$-cube), we have five bins which are centered at $\epsilon_r =$ 0.0371, 0.0458, 0.0659, 0.0859, and 0.1487. The overall mean dissipation-rate over the entire dataset is $\langle \epsilon \rangle = 0.093$ \cite{rf:Li08}.

To illustrate the situation, in Fig. \ref{Fig_Epsilon_r_contour_trajectory_128} we show 12 representative $64$-cubes placed inside the $1024^3$ domain, with 50 sample fluid particle trajectories emanating from each and progressing during a time equal to 27$\tau_K$. The required averages are taken over all these trajectories as well as over several cubes for which $\epsilon_r$ is in a bin's prescribed range.  The PDF of the dissipation-rate is shown in the insert in Fig. \ref{Fig_Epsilon_r_contour_trajectory_128} together with the bins or $\epsilon_r$ values used for the conditional averaging.

\begin{figure}[htbp]
   \begin{center}
      \includegraphics[width=2.7 in] {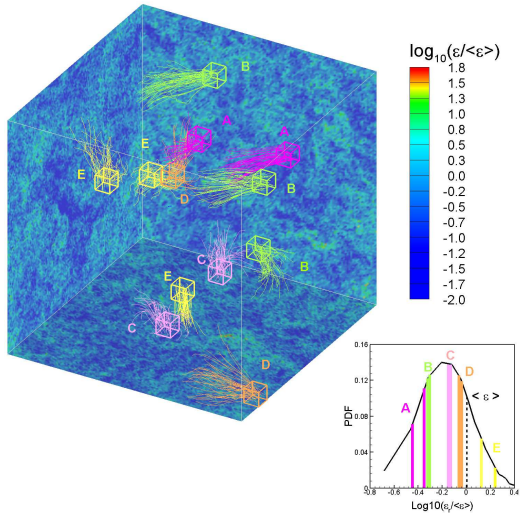}
      \caption{Sample particle trajectories starting from 12 randomly selected $64$-cubes characterized by local dissipation-rate $\epsilon_r$ at the initial time. Contours on the background planes is local (normalized) dissipation-rate (at the initial time) in logarithmic units showing its intermittent but structured distribution at the smallest scales of the flow. The lower right insert shows the PDF of the dissipation-rate together with the bins or $\epsilon_r$ values used for the conditional averaging.}
      \label{Fig_Epsilon_r_contour_trajectory_128}
   \end{center}
\end{figure}

For the case of $r=136\eta_K$ (64-cube), the various lines in Figure \ref{Fig_Epsilonsort_SO_tensor}(a) show the conditional auto-correlation functions so computed, plotted as function of time-delay scaled by global Kolmogorov time scale $\tau_K$. Averages are evaluated over $6 \times 10^3$ particles in each bin. Tests varying the number of particles show that the auto-correlation functions are well converged with this number of particles. There are noticeable differences in the results depending on $\epsilon_r$, for both the ${\bf S}$ and ${\bf \Omega}$ auto-correlation functions. For larger values of $\epsilon_r$, i.e. in regions of more intense turbulence activity, the decay rate of the auto-correlation functions is faster.  This shows that the global Kolmogorov time-scale $\tau_K = (\nu/\langle \epsilon \rangle)^{1/2}$ does not determine the {\it local} evolution of patches of turbulence, even in a statistical sense when conditional averaging is used that segregates different types of fluid regions.

As a next step, consistent with the RKSH, we define a `local Kolmogorov time-scale' based on the local dissipation-rate according to
\begin{equation}
\tau_{K,r} = (\nu/  \epsilon_r )^{1/2}.
\label{eq:taulocal}
\end{equation}

A Lagrangian version of the RKSH would state that the temporal auto-correlation functions should be a universal function of a time-delay $\tau$ normalized by a `local Kolmogorov time scale' $\tau_{K,r} $. Results shown in Fig. \ref{Fig_Epsilonsort_SO_tensor}(b) thus scaled by the local time-scale for each tensor show excellent collapse. In this plot we also show the auto-correlation for the velocity-gradient tensor ${\bf A}$ which falls between the results of its symmetric and antisymmetric parts. Again, the auto-correlation of the rotation-rate tensor decays much more slowly than that of the strain-rate tensor, which can be attributed to the effect of the deviatoric part of the pressure Hessian as discussed more in depth in Ref. \cite{rf:Yu09}. The deviatoric part of the pressure Hessian causes non-local effects on the strain-rate evolution but has no effects on the evolution of rotation-rate. Figure \ref{Fig_Epsilonsort_dp} shows similar results for the pressure gradient. Also here, the collapse is much more improved when using the local time-scale that corresponds to the local dissipation-rate, showing that the Lagrangian KRSH works for both the velocity and pressure gradients, the latter corresponding closely to the fluid acceleration \cite{rf:Vedula99,rf:Liu06}.

We point out that in recent work \cite{rf:Benzi08} the Lagrangian KRSH has also been shown to hold in the context of moments of two-time velocity increments. In their analysis, the authors use a rate of dissipation  $\epsilon_{\tau}$ averaged over {\it temporal} domains of duration $\tau$ along the particle trajectory.  The `fully Lagrangian' quantity $\epsilon_{\tau}$ averages, and thus connects, dissipation at various times. From a different, perhaps complementary point of view, the present analysis based on the more traditional spatial average of dissipation at a single (initial-condition) time facilitates interpreting the results in terms of `causality' based on a fixed initial condition.

\begin{figure}[htbp]
   \begin{center}
      \includegraphics[width=3.2 in] {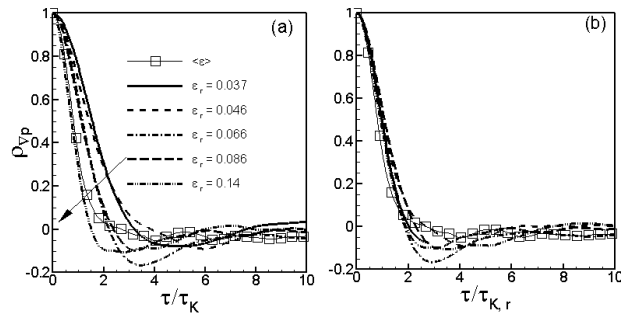}
      \caption{Lagrangian auto-correlations of pressure gradient. time-lag is normalized using (a) global time-scale $\tau_K$ and (b) local time scale $\tau_{K,r}$.}
      \label{Fig_Epsilonsort_dp}
   \end{center}
\end{figure}

Next, we explore the validity of the Lagrangian KRSH for different box sizes $r$.  Figure \ref{Fig_Epsilonsort_Sallbins_tensor} shows the auto-correlations of strain-rate tensor $\rho_{\bf S}$ for the four box sizes $r$ and $\epsilon_r$ described above. Now the scatter is even more pronounced in (a) where time-lag is normalized by the global averaged Kolmogorov time scale $\tau_K$, mainly due to the fact that the range of values of $\epsilon_r$ for smaller values of $r$ is larger due to intermittency. Once again, excellent collapse can be observed in (b) where the time-lag is normalized by the local Kolmogorov time scales $\tau_{K,r}$, even though the curves correspond to different $r$ and $\epsilon_r$ values. This is confirmed further by comparing the results with similar local $\epsilon_r$ values but different $r$:  e.g. the first two curves from the right to left shown in Fig. \ref{Fig_Epsilonsort_Sallbins_tensor} (a) correspond to $\epsilon_r=0.0076$ for a $16$-cube and $\epsilon_r=0.0096$ fora $32$-cube and they collapse quite well. Similar
results (not shown) are obtained for rotation rate tensor.

 \begin{figure}[htbp]
   \begin{center}
      \includegraphics[width=3.2 in] {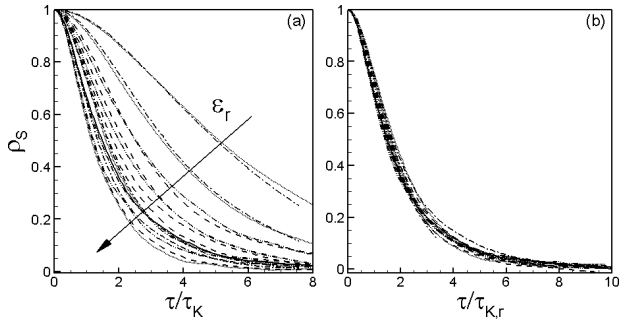}
      \caption{Lagrangian auto-correlations of strain-rate tensor computed from Eq. (\ref{eq:correlationfunction}) for four box-size values in the inertial range $r$, and for four $\epsilon_r$ values (except for $64$-cube which has 5 bins of $\epsilon_r$). The lines are solid lines : $16$-cube; $\cdots$ : $32$-cube; $--$: $64$-cube; $-\cdot \cdot$ : $128$-cube; open squares: global average. In (a) time is scaled using global time-scale, $\tau_K$, in (b) using the local time scale, $\tau_{K,r}$.}
      \label{Fig_Epsilonsort_Sallbins_tensor}
   \end{center}
\end{figure}

Present results provide a new and more dynamical interpretation of the Kolmogorov Similarity Hypothesis: when focussing on particular subregions of turbulence, the dynamics proceed according to time-scales dictated by the locally averaged rate of dissipation. The classical view of the KRSH focussed on the moments of velocity increments in turbulent fields that had already developed due to past dynamics. Here we have shown that  the dynamical time evolution of turbulence, its rate of change and associated decorrelation time scale, depends on the local rate of dissipation, averaged over a volume comparable to the volume containing the sample of initial particle locations. Results also suggest that Lagrangian models (e.g. for the velocity gradient tensor \cite{rf:Chevillard06} or the acceleration \cite{rf:Lamorgese07}) that require specification of a characteristic time scale should not use the global, but the local time scale.

We thank the Keck Foundation for postdoctoral support (HY), the National Science Foundation (ITR-0428325 and CDI-0941530) for its support of the public database and G. L. Eyink for insightful comments.

\end{document}